\documentclass[showpacs,amsmath,amssymb,twocolumn]{revtex4}
\usepackage{graphicx}
\usepackage{dcolumn}
\usepackage{bm}
\usepackage{overpic}

\begin{document}

\title{\bf\large{Phase Diagram of Neutron-Proton Condensate in Asymmetric Nuclear Matter}}
\author{\normalsize{Meng Jin$^{1,2}$, Lianyi He$^1$, and Pengfei Zhuang$^1$}}
\affiliation{$^1$Physics Department, Tsinghua University, Beijing
100084, China\\
$^2$Institute of Particle Physics, Central China Normal
University, Wuhan 430070, China}

\begin{abstract}
We investigate the phase structure of homogeneous and
inhomogeneous neutron-proton condensate in isospin asymmetric
nuclear matter. At extremely low nuclear density the condensed
matter is in homogeneous phase at any temperature, while in
general case it is in Larkin-Ovchinnikov-Fulde -Ferrell phase at
low temperature. In comparison with the homogeneous superfluid,
the inhomogeneous superfluid can survive at higher nuclear density
and higher isospin asymmetry.
\end{abstract}


\maketitle

It is well-known that the neutron-proton ($np$) pairing plays an
important role in nuclear physics and astrophysics, such as the
structure of medium-mass nuclei produced in radioactive nuclear
beam facilities\cite{goodman}, the deuteron formation in
intermediate energy heavy-ion collisions\cite{baldo}, the pion and
kaon condensation\cite{brown}, the $r$-process\cite{kratz,chen},
and the cooling of neutron stars. The microscopic calculations
show that the nuclear matter supports $np$ Cooper pairing in the
$^3$S$_1-^3$D$_1$ channel due to the tensor component of the
nuclear force, and the pairing gap is of the order of $10$
MeV\cite{alm,von,baldo2,alm2,alm3,sedrakian,elg} at the saturation
nuclear density. At low enough density the $np$ Cooper pairs would
go over to Bose-Einstein condensation(BEC) of deuterons in
symmetric nuclear matter\cite{alm2,baldo}.

The emergence of isospin asymmetry will generally suppress the
$np$ pairing, and the condensate will disappear when the asymmetry
becomes sufficiently large. Near the saturation density, the $np$
pairing correlation depends crucially on the mismatch between the
two Fermi surfaces, and a small isospin asymmetry can break the
condensate due to the Pauli blocking effect. At very low density,
when neutrons and protons start to form deuterons and when the
spatial separation between deuterons and between deuterons and
neutrons is large, the Pauli blocking loses its efficiency in
destroying a $np$ condensate. In such situation, the isospin
asymmetry can be very large, and the $np$ condensate survives in
the form of deuteron-neutron mixture in momentum
space\cite{lombardo,isayev}. Different from the symmetric nuclear
matter where the thermal motion destroys the $np$ condensate, for
asymmetric nuclear matter the temperature effect will melt the
condensate on one hand and increase the overlapping between the
two effective Fermi surfaces on the other hand. As a result of the
competition, in a wide density regime the temperature dependence
of the superfluidity is very strange\cite{sedrakian2,akhiezer}:
The maximum condensate is not located at zero temperature, and the
pairing even occurs only at intermediate temperature for large
isospin asymmetry.

The above results are obtained by assuming the condensate is
homogeneous in the ground state. What is the true phase structure
of $np$ condensate with isospin asymmetry when the inhomogeneous
Larkin-Ovchinnikov-Fulde -Ferrell(LOFF) phase\cite{LOFF} is taken
into account? How will the LOFF phase change the strange
temperature behavior of $np$ condensate found in
\cite{sedrakian2}? In fact, there should exist a rich phase
structure in asymmetric nuclear matter, since the isospin
asymmetry essentially plays the same role as the population
imbalance in the two-component resonantly interacting atomic Fermi
gas\cite{pao}. Different to the cold atoms , in nuclear matter the
phase separation at large length scale may be forbidden and the
LOFF phase may be energetically favored. Although the LOFF phase
was discussed in asymmetric nuclear matter at saturation
density\cite{LOFFN}, the phase structure including the LOFF phase
in the whole density, temperature and isospin asymmetry space and
the effect of the LOFF phase on the strange temperature behavior
of the $np$ condensate are still unknown. We will present in this
paper the phase diagrams in the density, temperature and isospin
asymmetry space.

The often used formulae for superfluid in nuclear matter are
discussed in detail in \cite{akhiezer2} where the superfluid state
is described by a normal and an anomalous nucleon distribution
functions ${\cal F}$ and ${\cal G}$. Generally, they are functions
of momentum ${\bf p}$ and matrices in spin and isospin space. The
formulae can be easily generalized to study the isospin asymmetric
superfluid with total pair momentum $2{\bf q}$. We will start with
the LOFF phase, and the homogeneous phase can be recovered by
taking ${\bf q}=0$. Like the studies in
\cite{sedrakian2,akhiezer,lombardo,LOFFN,isayev}, we discuss the
$np$ pairing in the $^3$S$_1-^3$D$_1$ channel with total spin
$S=1$, isospin $T=0$ and their projections $S_z=T_z=0$. In this
case the distribution functions take the structure
\begin{eqnarray}
{\cal F}({\bf p})&=&{\cal F}_{00}({\bf p})\sigma_0\tau_0+{\cal
F}_{03}({\bf
p})\sigma_0\tau_3,\nonumber\\
{\cal G}({\bf p})&=&{\cal G}_{30}({\bf p})\sigma_3\sigma_2\tau_2,
\end{eqnarray}
where $\sigma_i$ and $\tau_i$ are the Pauli matrices in spin and
isospin spaces. Using the minimum principle of the thermodynamic
potential and the procedure of block
diagonalization\cite{akhiezer3}, we can express the elements as
\begin{eqnarray}
{\cal F}_{00}({\bf p})&=&1/2-\xi_{\bf p}\left[1-f(E_{\bf p}^+)-f(E_{\bf p}^-)\right]/\left(2E_{\bf p}\right),\nonumber\\
{\cal F}_{03}({\bf p})&=&\left[f(E_{\bf p}^-)-f(E_{\bf p}^+)\right]/2,\nonumber\\
{\cal G}_{30}({\bf p})&=&-\Delta_{\bf p}\left[1-f(E_{\bf
p}^+)-f(E_{\bf p}^-)\right]/\left(2E_{\bf p}\right)
\end{eqnarray}
with the notations $\xi_{\bf p} = \left({\bf p}^2+{\bf
q}^2\right)/\left(2m\right)-\mu$, $E_{\bf p} = \sqrt{\xi_{\bf
p}^2+\Delta^2_{\bf p}}$ and $E_{\bf p}^\pm = E_{\bf
p}\pm\left(\delta\mu+{\bf p}\cdot{\bf q}/m\right)$, where $m$ is
the effective nucleon mass in the medium and $f(x)=1/(e^{x/T}+1)$
is the Fermi-Dirac function with $T$ being the temperature. We
have introduced the average chemical potential
$\mu=(\mu_n+\mu_p)/2$ and the mismatch $\delta\mu=(\mu_n-\mu_p)/2$
instead of the neutron and proton chemical potentials $\mu_n$ and
$\mu_p$. We have also neglected the possible neutron-proton mass
splitting induced by the isospin asymmetry which is believed to be
small. The $np$ condensate $\Delta_{\bf p}$ is generally momentum
dependent and satisfies the gap equation
\begin{equation}
\label{gap1} \Delta_{\bf p}=\int\frac{d^3{\bf k}}{(2\pi)^3}V({\bf
p},{\bf k}){\cal G}_{30}({\bf k}),
\end{equation}
where $V$ is the nucleon-nucleon (NN) interaction potential. The
LOFF momentum ${\bf q}$ should be determined via minimizing the
the free energy ${\cal E}$, which ensures the total current ${\bf
j}_s$ in the ground state to be zero,
\begin{eqnarray}
\label{gap2}\rho |{\bf q}|- 4\int\frac{d^3{\bf
p}}{(2\pi)^3}\frac{{\bf p}\cdot{\bf q}}{|{\bf q}|}{\cal
F}_{03}({\bf p})=0,
\end{eqnarray}
where we have used the total nucleon density $\rho=\rho_n+\rho_p$
and the isospin density asymmetry $\delta\rho=\rho_n-\rho_p$
instead of the neutron and proton densities $\rho_n$ and $\rho_p$,
\begin{equation}
\label{number} \rho=4\int\frac{d^3{\bf p}}{(2\pi)^3}{\cal
F}_{00}({\bf p}),\ \ \delta\rho=4\int\frac{d^3{\bf
p}}{(2\pi)^3}{\cal F}_{03}({\bf p}).
\end{equation}

Once the NN potential $V$ is known, we can solve the coupled set
of gap equations (\ref{gap1}) and (\ref{gap2}) together with the
density equations (\ref{number}) at given temperature $T$, baryon
density $\rho$ or equivalently the Fermi momentum
$k_F=(1.5\pi^2\rho)^{1/3}$ and isospin asymmetry
$\alpha=\delta\rho/\rho$, and obtain all possible phases, namely
the normal phase $\Delta_{\bf p}=0$, the homogeneous superfluid
phase $\Delta_{\bf p}\neq 0,\ {\bf q}=0$ and the LOFF phase
$\Delta_{\bf p}\neq 0,\ {\bf q}\neq 0$. By comparing their free
energies we can determine the true ground state.

The details of the phase diagram depend on the NN potential $V$ we
will chose in the numerical calculations, however, the qualitative
topology structure of the phase diagram does not depend on that.
To show this, we analyze the stability of the homogeneous
superfluid phase against the formation of a nonzero Cooper pair
momentum. For this purpose, we investigate the response of the
free energy ${\cal E}$ to a small pair momentum ${\bf q}$ via the
small ${\bf q}$ expansion, ${\cal E}({\bf q})={\cal E}({\bf
0})+{\bf j}_s\cdot{\bf q}/m+\rho_s{\bf q}^2/(2m)+\cdots$, where
${\bf j}_s=m\partial{\cal E}/\partial{\bf q}$ is the total current
which is proportional to the left hand side of (\ref{gap2}), and
$\rho_s$ is just the superfluid density defined by $\rho_s=
m\partial^2{\cal E}/\partial {\bf q}^2$ of which the explicit form
reads
\begin{equation}
\label{rhos} \rho_s=\rho+\frac{2}{m}\int\frac{d^3{\bf
p}}{(2\pi)^3}\frac{{\bf p}^2}{3}\left[f^\prime(E_{\bf
p}^+)+f^\prime(E_{\bf p}^-)\right]\big|_{{\bf q}=0}
\end{equation}
with the definition $f'(x)=df(x)/dx$. The current ${\bf j}_s$
vanishes due to the gap equation for ${\bf q}$, and the sign of
$\rho_s$ controls the stability of the homogeneous superfluid,
i.e., a negative $\rho_s$ means the LOFF phase has lower free
energy than the homogeneous superfluid phase.

The momentum dependence of the gap function $\Delta_{\bf p}$ is
normally rather weak in a wide momentum region\cite{lombardo}, and
we can approximately treat it as a constant $\Delta$ for a
qualitative analysis. At zero temperature, once the isospin
asymmetry is turned on, there must exist a sharp breached region
where the quasiparticle energy $E_{\bf p}^-<0$ with the necessary
condition $\delta\mu>\Delta$, and the momentum integration in
(\ref{rhos}) can be analytically carried out,
\begin{equation}
\rho_s=\rho\left[1-\frac{p_+^3\Theta(\mu_+)+p_-^3\Theta(\mu_-)}{3\pi^2\rho}\frac{\delta\mu}{\sqrt{\delta\mu^2-\Delta^2}}\right],
\end{equation}
where $p_\pm=\sqrt{2m\mu_\pm}$ are possible gapless nodes with
$\mu_\pm=\mu\pm \sqrt{\delta\mu^2-\Delta^2}$. At high density the
matter is in BCS regime where $\delta\mu,\Delta\ll\mu$, and the
breached region is $p_-<|{\bf p}|<p_+$. Since $p_\pm$ are close to
the Fermi momentum $k_F$, the superfluid density should be
negative since $\rho_s\simeq
\rho(1-\delta\mu/\sqrt{\delta\mu^2-\Delta^2})$. On the other hand,
at low enough density the matter is in BEC regime, the chemical
potential $\mu$ becomes negative which leads to $\mu_-<0$ and a
reduced breached region $0<|{\bf p}|<p_+$. In this case $p_+$ is
much smaller than $k_F$ and the superfluid density becomes
positive. Therefore, at zero temperature the superfluid is
expected to evolve from an inhomogeneous phase to the homogeneous
phase when the nuclear density decreases, which is a general
phenomenon for BCS-BEC crossover with population
imbalance\cite{pao}.

At finite temperature, the breached region is smeared due to the
thermal excitation, and $\delta\mu>\Delta$ is not necessary. At
the critical temperature $T_c$ there should be a second order
phase transition from the superfluid to normal state, and for
temperature $T\lesssim T_c$ the pairing gap behaves as
$\Delta(T)\propto(1-T/T_c)^{1/2}$ which leads to the regular
behavior of the superfluid density $\rho_s(T)\propto(1-T/T_c)>0$.
This means the temperature tends to stabilize the homogeneous
phase. Combining with the behavior of the superfluid density at
zero temperature, the homogeneous phase at sufficiently low
density will keep stable at any temperature below $T_c$, while at
high density there must exist a turning temperature $T_s$ where
$\rho_s$ changes sign, and the superfluid should be in
inhomogeneous phase at low temperature $T<T_s$ and in homogeneous
phase at high temperature $T_s<T<T_c$. Since the homogeneous state
is unstable at low temperature, combining with the fact that the
critical isospin asymmetry for the LOFF phase is much larger than
the homogeneous phase\cite{LOFFN}, the strange temperature
behavior of the pairing gap found in \cite{sedrakian2} is probably
unrealistic.
\begin{figure}
\centering
\includegraphics[width=7cm]{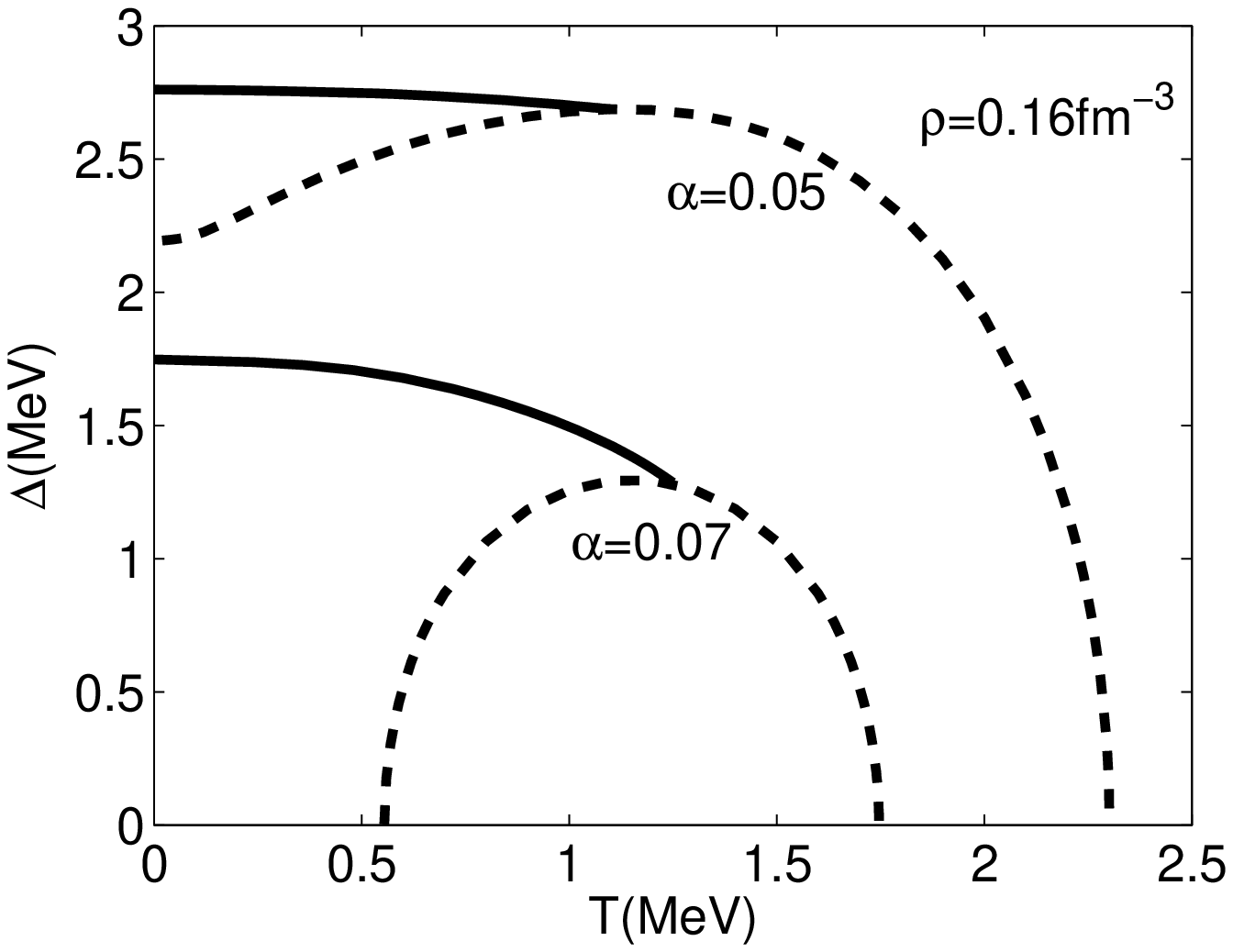}
\includegraphics[width=7cm]{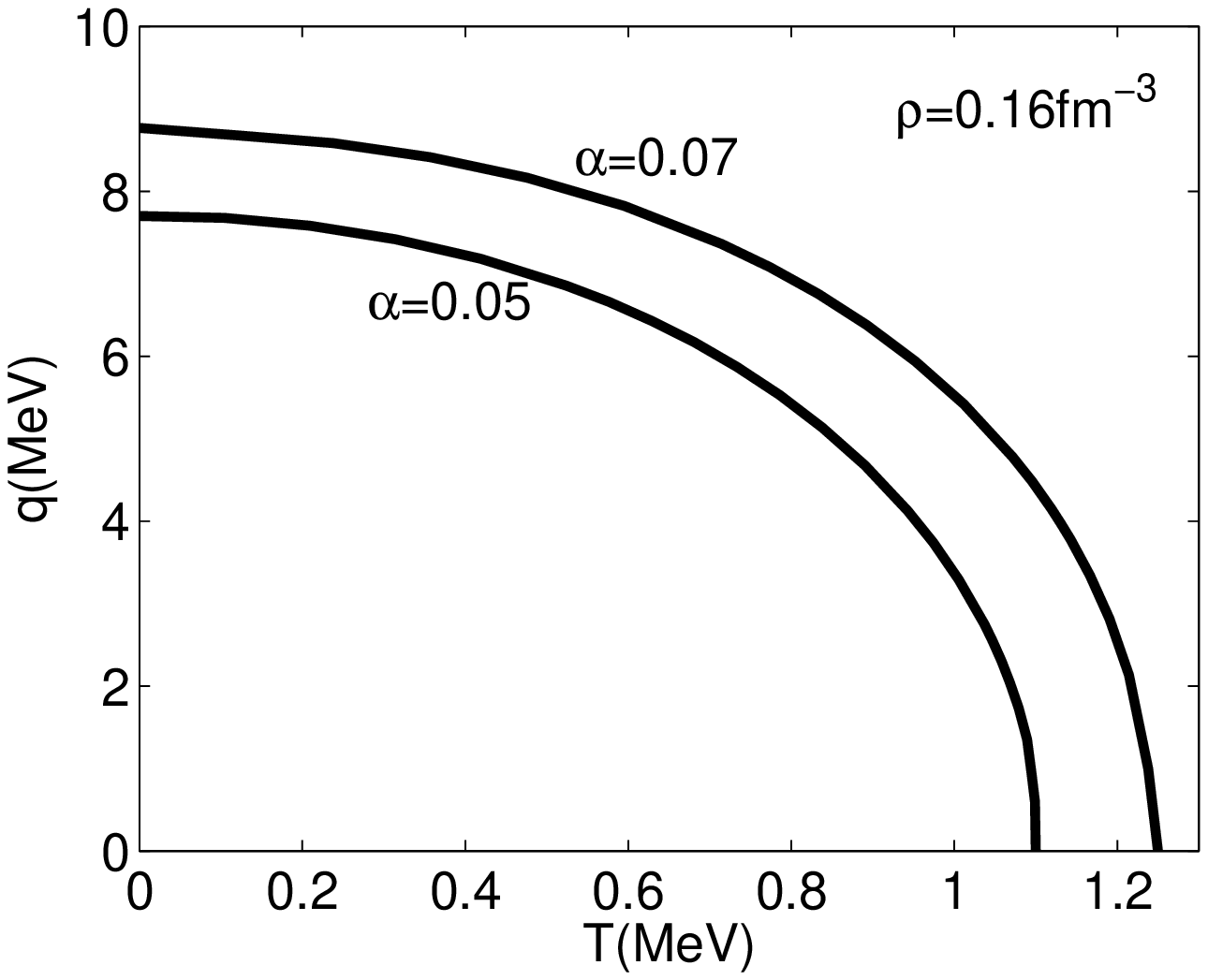}
\caption{The pairing gap $\Delta$ for homogeneous (dashed lines)
and inhomogeneous (solid lines) condensates and LOFF momentum $q$
as functions of temperature at normal density $\rho_0$ and for two
values of isospin asymmetry. \label{fig1}}
\end{figure}

\begin{figure}
\centering
\includegraphics[width=6.6cm]{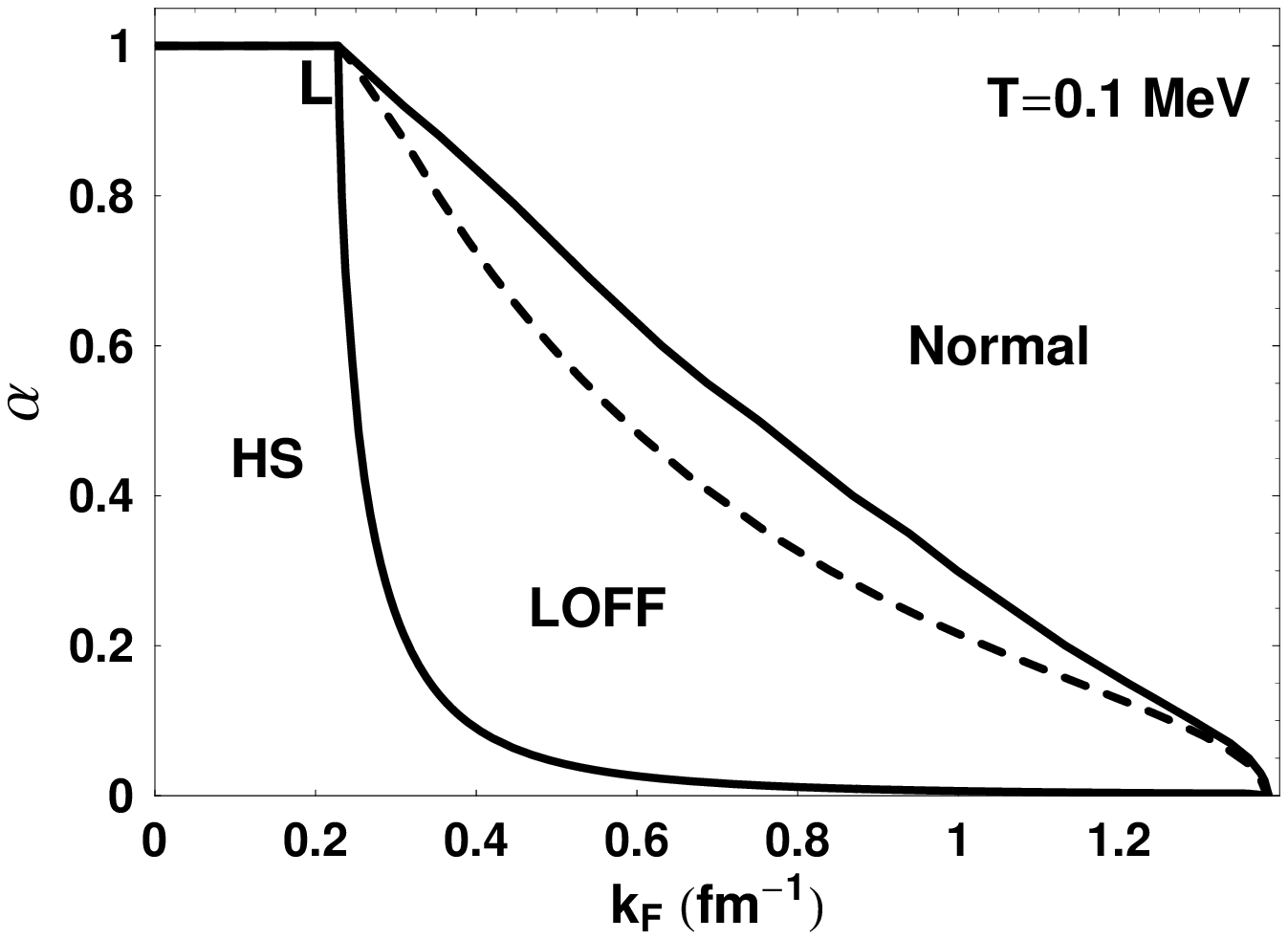}
\includegraphics[width=7cm]{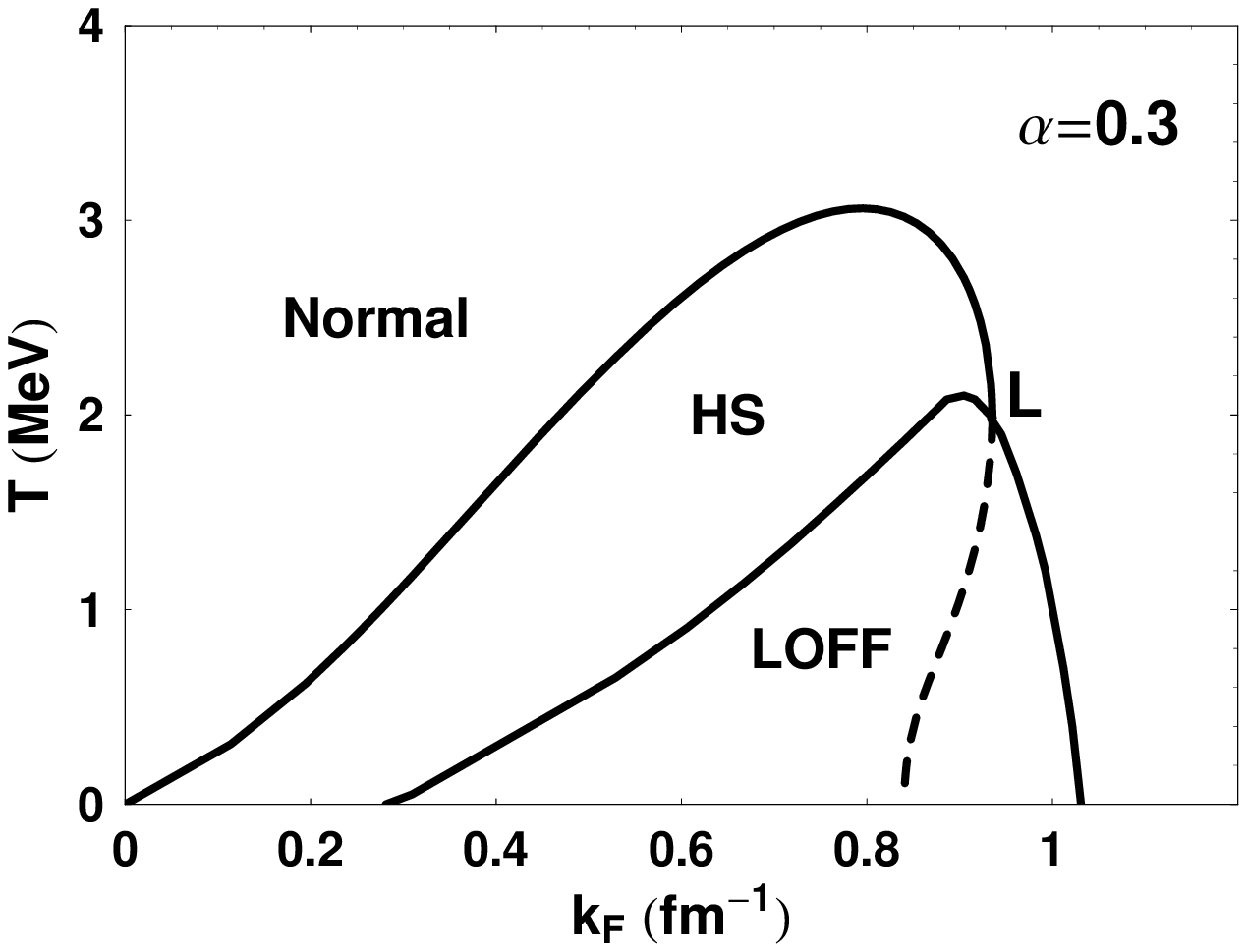}
\includegraphics[width=7cm]{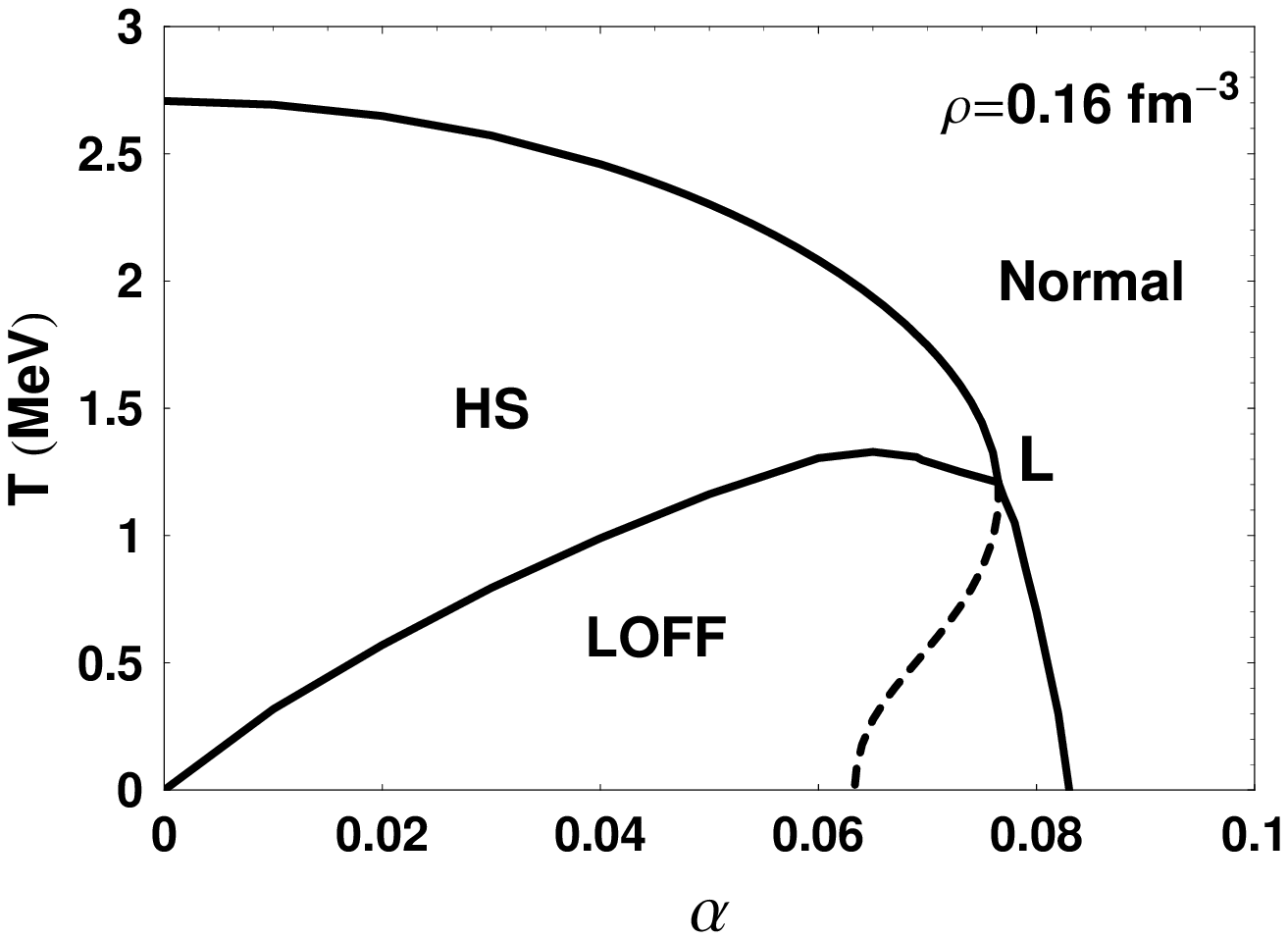}
\caption{The phase diagrams in $k_F-\alpha$, $k_F-T$ and
$\alpha-T$ planes. In each plane, the labels HS, LOFF and Normal
indicate homogeneous superfluid, LOFF superfluid and normal phase,
the dashed line is the border of the unstable HS, and the three
phases meet at a Lifshitz point L. \label{fig2}}
\end{figure}

We now move to numerical calculations. The Paris NN potential is
often used to describe the nuclear structure and nucleon
superfluidity, and describes well the BCS-BEC crossover of $np$
condensate\cite{baldo,lombardo}. Since the qualitative topology
structure of the phase diagram does not depend on specific models,
for the sake of simplicity, we employ a separable form of the
Paris NN potential
\begin{equation}
V({\bf r}_1,{\bf r}_2)=v_0\left[1-\eta\left({\rho\left(\frac{{\bf
r}_1+{\bf r}_2}{2}\right)\over
\rho_0}\right)^\gamma\right]\delta({\bf r}_1-{\bf r}_2),
\end{equation}
which was developed in \cite{garrido} to reproduce the pairing gap
in $S=1,T=0$ channel and the bound state between zero energy and
deuteron binding energy, where $\rho_0$ is the normal nuclear
density and the parameters $v_0, \eta, \gamma$ and an energy
cutoff $\varepsilon_c$ to regulate the model are determined by
recovering the pairing gap in the realistic Paris NN potential.
This separable form can also describe well the BCS-BEC crossover
of $np$ condensate\cite{isayev}. In our numerical calculation, we
choose $v_0=-530$ MeV$\cdot$fm$^3$, $\eta=0, \varepsilon_c=60$ MeV
and take $m$ as the density-dependent nucleon mass corresponding
to the Gogny force D1S\cite{d1s}. We have checked that different
parameter sets\cite{garrido} lead to only a slight change in the
numerical results.

In the low density BEC regime, the homogeneous phase is stable at
any temperature below $T_c$, and the condensate is a regular
decreasing function of temperature. Beyond this regime the strange
phenomenon of the homogeneous condensate arises: The maximum
pairing gap is located at non-zero temperature, and the condensate
even appears only at intermediate temperature with two critical
temperatures $T_o$ and $T_c$ for large isospin asymmetry, which is
shown as dashed lines in the upper panel of Fig.\ref{fig1} for
saturation density without loss of generality. However, the
superfluid density $\rho_s$ for the homogeneous phase is negative
at $T<T_s$ and positive at $T_s<T<T_c$, where the turning
temperature $T_s$ is larger than the lower critical temperature
$T_o$. This tells us that the homogeneous phase is stable at high
temperature $T>T_s$ and unstable to formation of LOFF condensate
at low temperature $T<T_s$. By calculating the LOFF pairing gap
$\Delta$ and momentum $q=|{\bf q}|$ which are shown as solid lines
in the upper and lower panels of Fig.\ref{fig1} and comparing the
free energies for the homogeneous and LOFF states, the LOFF phase
is energetically more favored than the homogeneous phase at
$T<T_s$. Especially, different to the homogeneous condensate, the
LOFF condensate always starts at zero temperature. Therefore,
after considering both homogeneous and inhomogeneous condensates,
the strange temperature dependence of the pairing
gap\cite{sedrakian2,akhiezer} disappears and the condensate
becomes a regular decreasing function of temperature. The LOFF
momentum $q$, shown in the lower panel of Fig.\ref{fig1}, drops
down with increasing temperature and approaches to zero
continuously at $T_s$, which indicates a continuous phase
transition from homogeneous phase to LOFF phase. The continuity
can be proven analytically. The gap equation (\ref{gap2}) for $q$
can be written as $qW(q)=0$ with a trivial solution $q=0$ for the
homogeneous phase and a non-zero solution from $W(q)=0$ for the
LOFF phase. Using the expansion for ${\cal E}$ we find
$\rho_s=W(0)$. Therefore, we must have $q=0$ at $T=T_s$, providing
that the LOFF solution is unique.

The phase diagrams of the $np$ pairing are shown in
Fig.\ref{fig2}. We first discuss the one in the $k_F-\alpha$ plane
at a very low temperature $T=0.1$ MeV. When $\rho\rightarrow 0$ we
find $\mu\rightarrow-\varepsilon_b/2$ at $\alpha=0$ where
$\varepsilon_b$ is the deuteron binding energy, which means that
the $np$ condensate survives in the form of deuteron BEC.
Consistent with the findings in atomic Fermi gas\cite{pao}, the
homogeneous phase (HS) is stable only at very low density BEC
regime. In this regime, the critical isospin asymmetry can be very
large, and even approaches to $1$ for $k_F<0.23\ $fm$^{-1}$.
Beyond this extremely low density regime, the superfluid density
of the homogeneous phase becomes negative which indicates that the
LOFF phase is energetically favored. By calculating the superfluid
density of the HS phase and the LOFF solution, we can determine
the phase boundaries between HS and LOFF and between LOFF and
normal phase. The LOFF momentum is large at high $\rho$ and high
$\alpha$ and approaches to zero at the HS-LOFF boundary which
means a continuous phase transition. If we consider HS only, the
HS-Normal boundary (dashed line) is below the LOFF-Normal
boundary, this shows that the introduction of LOFF phase enlarges
the superfluid region. In the $k_F-T$ and $\alpha-T$ planes, the
HS and LOFF phases are separated by the turning temperature $T_s$.
Note that $T_s$ starts at $k_F\ne 0$ in $k_F-T$ plane which
corresponds to the stable HS at extremely low density but starts
at $\alpha=0$ in $\alpha-T$ plane for high density which means
that the HS with small isospin asymmetry is easy to be stabilized.
Again, in comparison with the calculation with only HS, the
superfluid is extended to higher density or higher asymmetry
region due to the introduction of the LOFF phase, and the unstable
HS-Normal boundary (dashed lines) which reflects the strange
``intermediate temperature superfluidity" is replaced by the
LOFF-Normal boundary. The phase transitions in the three planes
are all of second order, and in any case the three phases meet at
a Lifshitz point L\cite{lifshitz}. The $\alpha-T$ phase diagram we
obtained is very similar to the generic phase diagram of
two-component ultracold Fermi gas in a potential
trap\cite{machida}.

In summary, we have qualitatively investigated the phase structure
of $np$ condensate in isospin asymmetric nuclear matter and
confirmed our analysis with a model NN potential. The important
findings are: 1) The LOFF phase is the ground state in a wide
region of nuclear density, temperature and isospin asymmetry,
except for very low density and high temperature. 2) The strange
temperature behavior of the $np$ condensate is washed out by the
LOFF phase at low temperatures. 3)The superfluid region is
expanded to high density and high asymmetry due to the
introduction of LOFF phase.

{\bf Acknowledgement:} The work was supported by the grants
NSFC10575058, 10428510, 10435080 and 10447122.

\end{document}